\def\ONE{1}

\def\SHOWCOMMENTS{1}
\def\EXTRAMARGINFORCOMMENTS{0}

\documentclass[12pt, letterpaper]{article}

\usepackage[textsize=small,color=lightgray!20]{todonotes}
\usepackage[dvipsnames]{xcolor}
\newcounter{commentCounter}
\newcommand{\comm}[3]{\ifx\SHOWCOMMENTS\ONE\refstepcounter{commentCounter}\todo[bordercolor=#1, backgroundcolor=#1!10, #2]{{\color{#1} \bf \arabic{commentCounter}.} #3}\fi}

\ifx\SHOWCOMMENTS\ONE
\ifx\EXTRAMARGINFORCOMMENTS\ONE
\def\extraMargin{6cm}
\def\papwid{215.9mm}
\def\paphei{279.4mm}
\paperwidth=\dimexpr \papwid + \extraMargin + \extraMargin\relax
\paperheight=\dimexpr \paphei + \extraMargin\relax
\oddsidemargin=\dimexpr\oddsidemargin + \extraMargin\relax
\evensidemargin=\dimexpr\evensidemargin + \extraMargin\relax
\marginparwidth=\dimexpr \marginparwidth + \extraMargin\relax
\fi
\fi

\usepackage{tikz}
\usepackage{subfig}
\usepackage{subcaption}
\usepackage{environ}
\NewEnviron{graph}[1][]{%
    \begin{tikzpicture}[#1, very thick,
                        vertex/.style={circle, draw, thick, minimum size=4ex, inner sep=0.5ex},
                        point/.style={minimum size=0ex, inner sep=0ex, outer sep=0ex}]
        \BODY
    \end{tikzpicture}%
}
\usepackage{tabularx}

\usepackage{enumitem}
\usepackage{parskip}
\usepackage{float}
\usepackage{newfloat}
\DeclareFloatingEnvironment[fileext=algo, listname={List of Algorithms}, name=Pseudocode, placement=tbhp]{algorithm}
\usepackage{caption}
\floatstyle{boxed}
\restylefloat{algorithm}

\usepackage{clrscode3e}
\usepackage{varwidth} 
\def\gets{\leftarrow}

\def\slit{\\[-2.5ex]}

\usepackage[utf8]{inputenc}
\usepackage{amsmath, amssymb, amsthm}
\usepackage[hidelinks]{hyperref}
\usepackage{cleveref}

\usepackage{tcolorbox}
\usepackage{soul}
\usepackage{makecell}
\usepackage{tikz}
\usetikzlibrary{decorations.pathreplacing}
\setul{2pt}{0.75pt}

\usepackage{algorithmicx}
\usepackage[noend]{algpseudocode}
\usepackage{graphicx}
\graphicspath{{deneme/}}

\usepackage{cite}

\newtheorem{theorem}{Theorem}[section]
\newtheorem{lemma}[theorem]{Lemma}
\newtheorem{corollary}[theorem]{Corollary}

\theoremstyle{definition}

\theoremstyle{remark}

\algdef{SE}[DOWHILE]{Do}{doWhile}{\algorithmicdo}[1]{\algorithmicwhile\ #1}%

\newcommand{\dfn}[1]{\textit{#1}} 
\newcommand{\fcs}[1]{\ul{#1}} 

\usepackage{braket}
\usepackage{xspace}
\renewcommand{\emptyset}{\ensuremath{\varnothing}}
\newcommand{\Reals}{\ensuremath{\mathbb{R}}}
\newcommand{\APSP}{\textsc{Apsp}\xspace}
\newcommand{\APLCA}{\textsc{Aplca}\xspace}
\newcommand{\LCA}{\ensuremath{LC\hspace{-0.3ex}A}}
\newcommand{\LCAij}[2]{\ensuremath{\LCA(#1, #2)}}
\newcommand{\DCA}{\ensuremath{DC\hspace{-0.3ex}A}}
\newcommand{\DCAij}[2]{\ensuremath{\DCA(#1, #2)}}
\newcommand{\clo}[0]{clo}
\newcommand{\red}[0]{red}
\newcommand{\True}[0]{\ensuremath{T}}
\newcommand{\False}[0]{\ensuremath{F}}
\newcommand{\Eclo}[0]{\ensuremath{E_{\clo}}}
\newcommand{\Eess}[0]{\ensuremath{E^*}}
\newcommand{\Ered}[0]{\ensuremath{E_{\red}}}
\newcommand{\Eij}[2]{\ensuremath{E_{#1#2}}}
\newcommand{\Gess}[0]{\ensuremath{G^*}}
\newcommand{\Gred}[0]{\ensuremath{G_{\red}}}
\newcommand{\Gessij}[2]{\ensuremath{\Gess_{#1#2}}}
\newcommand{\Gij}[2]{\ensuremath{G_{#1#2}}}
\newcommand{\depth}{\ensuremath{depth}}
\newcommand{\depthi}[1]{\depth(#1)}
\newcommand{\IN}{\ensuremath{\textsc{in}}}
\newcommand{\INess}{\ensuremath{\IN^*}}
\newcommand{\INred}{\ensuremath{\IN_{\red}}}
\newcommand{\dmat}{\ensuremath{d}}

\newcommand{\INessv}[1]{\ensuremath{\INess(#1)}}
\newcommand{\INredv}[1]{\ensuremath{\INred(#1)}}
\newcommand{\INessij}[2]{\ensuremath{\INess_{#1#2}}}
\newcommand{\INessijv}[3]{\ensuremath{\INessij{#1}{#2}(#3)}}
\newcommand{\INv}[1]{\ensuremath{\IN(#1)}}
\newcommand{\DistIN}[0]{\ensuremath{D^*_{\IN}}}
\newcommand{\DistINt}[1]{\ensuremath{\DistIN}(#1)}
\newcommand{\DistINred}[0]{\ensuremath{D^{red}_{\IN}}}
\newcommand{\DistINredt}[1]{\ensuremath{\DistINred(#1)}}
\newcommand{\DistOUTred}[0]{\ensuremath{D^{red}_{\OUT}}}

\newcommand{\narset}[1]{\ensuremath{\set{\!#1\!}}}
\newcommand{\norm}[1]{\ensuremath{\left|#1\right|}}
\newcommand{\minfun}[1]{\ensuremath{\min\left(#1\right)}}

\newcommand{\OUT}{\ensuremath{\textsc{out}}}
\newcommand{\OUTess}{\ensuremath{\OUT^*}}
\newcommand{\OUTessij}[2]{\ensuremath{\OUTess_{#1#2}}}
\newcommand{\OUTessijv}[3]{\ensuremath{\OUTessij{#1}{#2}(#3)}}
\newcommand{\OUTijv}[3]{\ensuremath{\OUT_{#1#2}(#3)}}
\newcommand{\OUTessv}[1]{\ensuremath{\OUTess(#1)}}
\newcommand{\OUTv}[1]{\ensuremath{\OUT(#1)}}
\newcommand{\DistOUT}[0]{\ensuremath{D^*_{\OUT}}}
\newcommand{\DistOUTt}[1]{\ensuremath{\DistOUT}(#1)}
\newcommand{\such}{\ensuremath{\mid}}
\newcommand{\Vij}[2]{\ensuremath{V_{#1#2}}}
\newcommand{\DeltaINt}[1]{\ensuremath{\Delta_{\IN}(#1)}}
\newcommand{\DeltaINred}[0]{\ensuremath{\Delta_{\IN}}}
\newcommand{\DeltaINredt}[1]{\ensuremath{\DeltaINred(#1)}}
\newcommand{\DeltaOUTt}[1]{\ensuremath{\Delta_{\OUT}(#1)}}
\newcommand{\basicterm}{\ensuremath{V\!\Eess}}
\newcommand{\basictermE}{\ensuremath{V\!E}}
\newcommand{\refinedterm}{\ensuremath{(\Eess)^2/V}}
\newcommand{\refinedtermE}{\ensuremath{(E)^2/V}}
\newcommand{\refinedtermEred}{\ensuremath{\Ered^2/V}}
\newcommand{\covariantterm}{\ensuremath{V\sigma}}
\newcommand{\covarianttermred}{\ensuremath{V\sigma}}
\newcommand{\sigmasquareINred}{\ensuremath{\sigma^2_{\IN}}}
\newcommand{\sigmasquareOUTred}{\ensuremath{\sigma^2_{\OUT}}}
\newcommand{\variantterm}{\ensuremath{V\sigmasquareINred}}
\newcommand{\variantmaxterm}{\ensuremath{V(\sigmasquareINred + \sigmasquareOUTred)}}
\newcommand{\variantmixterm}{\ensuremath{V(\sigmasquareINred + \sigma)}}
\newcommand{\quadraticterm}{\ensuremath{V^2}}
\newcommand{\sigmared}{\ensuremath{\sigma}}
\newcommand{\muV}{\ensuremath{\mu{V}}}
    
\begin{document}
	
	\title{New Parameterized Upper Bounds for All-Pairs Shortest Path and Lowest Common Ancestor in
           DAGs\footnote{This manuscript is a \ul{preliminary and reduced} presentation of the work. The authors expect to
           publish an extended and detailed manuscript in the future. The reader is encouraged to check for newer
           versions of the manuscript.}}
	\author{Ersel Hengirmen\\Dept. of Computer Engr.\\Middle East Tech. Univ.\\Ankara, Turkey\\\texttt{erselh@metu.edu.tr}\and
	        Hakan Y{\i}ld{\i}z\\Dept. of Computer Engr.\\Middle East Tech. Univ.\\Ankara, Turkey\\\texttt{hakan@metu.edu.tr}}
	\date{}
	\maketitle

	\begin{abstract}
	    We present new algorithms on two well-known all-pairs problems on directed acyclic graphs (DAGs): shortest path and
lowest common ancestor. The new algorithms provide specialized bounds that relate to the degree distribution of the
vertices on the edges essential for the output. In particular, we show how to solve all-pairs shortest path problem on
real-weighted DAGs in $O(\refinedterm + \covariantterm + \quadraticterm)$ time, where $\Eess$ is the number of essential
edges and $\sigma$ is the covariance of the in-degree and out-degree exceedance distributions on these edges. In
comparison, the previous best bound making use of $\Eess$ was $O(\basicterm)$, which can be asymptotically larger. The
lowest common ancestor problem on DAGs also gets a similar bound, involving the variance of the in-degree exceedances.
At their heart, our algorithms leverage the varying branching factors in multiple recurrences on the same function to
obtain a bound that involves the covariance/variance. This interesting technique can be utilized in other places where
dynamic programming is used.

	\end{abstract}
    
	
    \section{Introduction}
	Directed acyclic graphs (DAGs) are a fundamental form of graphs that naturally represent entities with dependencies and
frequently act as backbones of data structures. Many problems on these graphs can be solved faster than on general
directed graphs, leveraging the fact that their vertices can be topologically ordered. Well-known problems such as
single-source shortest paths, single-source path counting and minimum spanning arborescence can be trivially solved in
optimal $\Theta(V + E)$ time, thanks to the topological order which can be computed within the same bound.

When a graph problem contains an ``all-pairs'' notion, however, an optimal algorithm is typically not obvious and thus
there is room for improvement. All-pairs DAG problems generally have a more detailed history than their single-source
counterparts, with various algorithms respecting different parameters. Continuing this line of research, we here present
algorithms that work within new parameterized bounds for three fundamental problems on DAGs, namely, all-pairs shortest
paths, transitive closure/reduction, and all-pairs lowest common ancestor.

It is worth to immediately highlight a few aspects of our approach. Firstly, our algorithms aim for a bound specialized
in the ``essential'' portion of the graph, which are the edges critical for the output, i.e., the minimal subset of
edges that keeps the output of the problem intact. In the case of shortest paths, this subset is known as the essential
subgraph \cite{mcgeoch1995all}. In the case of transitive closure and lowest common ancestor, it is the transitive reduction
\cite{goralvcikova1979reduct}. Naturally, some graphs possess an essential subset that is significantly smaller than the full edge set.

A more important aspect, however, is that our methods utilize a parameter that can enable a more refined time bound
analysis compared to what is enabled by just using the essential subgraph size. While this parameter can take various
forms based on the problem, it essentially is a variance on the degree exceedance distribution of the vertices
restricted to the essential edges of the graph. The degree exceedance distribution is a function $D$ such that $D(t)$
equals the number of vertices whose degree (inclusively) exceeds $t$. (One can think of it as an inverted cumulative
degree distribution.) 

The technique we use to obtain the variance bound is particularly interesting. In rough terms, we greedily choose
among multiple recurrences defined on the same function to employ dynamic programming.\footnote{This is not to be
confused with the usual greedy algorithms that are based on a single recurrence.} This approach results
in a time bound that involves a covariance (or a variance) based on the possible costs of the recurrence.
The idea seems generalizable to other dynamic programming instances to obtain parameterized bounds.

In the light of the aspects mentioned above, we now summarize our results more precisely as follows:
\begin{enumerate}[label=(\alph*)]
    \item For real-weighted DAGs, we describe an algorithm that computes the all-pairs shortest paths in $O(\refinedterm
          + \covariantterm + \quadraticterm)$ time, where $V$ is the number of vertices, $\Eess$ is the size of the
          essential subgraph, and $\sigma$ is the covariance of the in-degree and out-degree exceedance distributions of
          the vertices in the essential subgraph. We show that this bound can be asymptotically better than the previous
          best bound that used $\Eess$, which is $O(\basicterm)$ \cite{chung2007essential}. As part of the discussion on this algorithm,
          we also review basic properties of the essential subgraph.
    \item The result in $(a)$ extends to the computation of the transitive closure and reduction on DAGs with the same
          time bound, with one minor exception. Due to the unweighted nature of the problem, the essential subgraph
          becomes the transitive reduction, whose size we represent with $\Ered$ (rather than $\Eess$). Also, the
          covariance $\sigma$ is restricted to the transitive reduction.
    \item We show how to compute a lowest common ancestor for each pair of vertices in a DAG in $O(\refinedtermEred +
          \variantmixterm + \quadraticterm)$ time, where $\sigma_{\IN}$ is the variance of the in-degree exceedances.
          Both $\sigma$ and $\sigma_{\IN}$ is restricted to the transitive reduction. (In the full version of the paper,
          we will also describe how to extend this result with a similar bound to the $k$-LCA problem, which is the
          lowest common ancestor problem for $k$-tuples rather than pairs.)
\end{enumerate}

The paper is organized as follows: \Cref{sec:related} describes related work from the literature. \Cref{sec:apsp}
explains the essential subgraph concept, followed by our all-pairs shortest path algorithm, which is
then extended to transitive closure/reduction. \Cref{sec:aplca} explains the all-pairs lowest common ancestor algorithm.

	\subsection{Related Work}
	\label{sec:related}
	\paragraph{Shortest paths.}
The basic form of the shortest path problem is the single-source version on general directed graphs with real weights,
which can be solved via textbook algorithms such as Bellman-Ford \cite{ford1956network,bellman1958routing} and Dijkstra
\cite{Dijkstra1959}, the latter achieving a time bound of $O(E + V\log{V})$ with Fibonacci heaps
\cite{fredman1987fibonacci}.\footnote{$V$ is the number of vertices and $E$ is the number of edges.} In DAGs, the
single-source problem becomes trivial and can be solved in $\Theta(V + E)$ time via topological sorting.

The all-pairs version of the problem on general real-weighted directed graphs is easily solved by invoking a
single-source algorithm $V$ times. Nonetheless, the usual textbook algorithm presented for the problem is the
$O(V^3)$-time Floyd-Warshall algorithm thanks to its elegance\cite{warshall1962theorem}. Over
the decades, the upper bound has steadily decreased \cite{madkour2017survey}, eventually reaching
$V^3/2^{\Theta(\sqrt{\log{V}})}$ thanks to an algorithm by Williams \cite{williams2014faster}. Also, Pettie's work
achieves $O(VE + V^2 \log\log{V})$, incorporating $E$ as a parameter \cite{pettie2004new}. The algorithms by Mcgeoch
\cite{mcgeoch1995all} and Karger et al. \cite{karger1993finding} are the most related to our work since they find
and use an underlying essential part of the graph in $O(VE^*+V^2\log V)$ time where $\Eess$ is the size of this
essential part. (We note that the essential parts used by each work are similar but
technically different. The essential subgraph we refer in this paper is the one used by \cite{mcgeoch1995all}. The
essential part in \cite{karger1993finding} can be larger; we have a further note on this later in this manuscript.) 

All-pairs shortest paths on DAGs can be solved using iterative application of single-source shortest path resulting
in $O(VE + V^2)$ time. Chung \cite{chung2007essential} provided a specialized upper bound of $O(VE^*+V^2)$ time, with
his algorithm using the same essential edges as Karger et al. \cite{karger1993finding}.\footnote{Chung's paper actually
claims a complexity of $O(VE^*+E)$ but that seems to ignore the initialization of the distance matrix. Also, a direct
interpretation of their pseudocode results in a $V^2$ term; this term can indeed be eliminated (for certain
output formats) but it is not explained in their paper.}

\paragraph{Transitive closure.} Computing the transitive closure of a relation, which is essentially a directed graph,
is an all-pairs problem by nature. The transitive closure can easily be computed by repeated application of a basic
graph search starting from each node in $O(VE+V^2)$ time. Fischer et al. was able to use matrix multiplication to
achieve an upper bound of $O(n^\omega)$ \cite{fischer1971boolean}. Some researchers focused on also finding the
transitive reduction of the graph, which is the essential part of the graph excluding of its edges with alternative
transitive paths. The term ``transitive reduction'' was coined by Aho et al.\cite{aho1972transitive}. Goralčíková and
Koubek showed how to compute the reduction in $O(VE_{red})$ time, where $E_{red}$ is the size of the reduction
\cite{goralvcikova1979reduct}. There also exists a line of work providing parametric bounds via chain decompositions and
covers \cite{simon1988improved, kogan2022beating}.

\paragraph{Lowest common ancestor.} Traditionally a concept defined in trees, the lowest common ancestor can also be
defined in DAGs, as a DAG also presents a clean ancestor-descendant relationship between its vertices. The lowest common
ancestors any two vertices can be found in can be trivially found in $O(V + E)$ time in DAGs.

The all-pairs version of the problem, where one finds a lowest common ancestor for every pair of vertices, is less
straightforward. Czumaj et al. introduced an algorithm that works in $O(VE)$ \cite{czumaj2007faster}. Later, this
bound was improved to $O(VE\log(\frac{V^2}{E})/\log{V})$ by Blelloch et al.\cite{blelloch2008new}. For dense
graphs, one of the most well-known algorithms is due to Czumaj et al., which achieved $O(V^{2.575})$ time using
algebraic methods \cite{czumaj2007faster}. A further improvement by Grandoni et al. \cite{grandoni2021all} reduced the
bound to $\widetilde{O}(V^{2.447})$, by using chain/antichain decomposition with fast matrix multiplication. There also
exists other research leveraging the chains in the graph \cite{kowaluk2007unique, kowaluk2008path}.

Out of all research, Eckhardt et al.'s is closest to our approach \cite{eckhardt2007fast}. As a first step, they extract
the transitive reduction via an algorithm by Simon \cite{simon1988improved}. This in turn allows them to solve all-pairs
within a parameterized time bound of $O(VE_{red})$ where $E_{red}$ is the size of the reduction.

    \section{All-Pairs Shortest Paths in DAGs}
    \label{sec:apsp}

    In this section, we describe our algorithm for the all-pairs shortest path problem on DAGs. The algorithm has a
    parameterized time bound based on a subset of the DAG known as its essential subgraph. The parameter in the bound
    involves a covariance on the directional degree exceedance distributions (essentially inverted cumulative
    degree distributions) of the vertices, restricted to the essential subgraph. We show that this new parameterized
    bound is better for some classes of graphs. Finally, we extend our result to transitive closure computation.

	\subsection{Preliminaries}
	\label{sec:preliminaries}
	Consider a directed acyclic graph $G$ with real weights given as a pair $(V, E)$ where $V$ is $G$'s set of vertices and
$E$ is $G$'s set of edges. (Let us quickly note that we abuse the symbols $V$ and $E$ to also refer to the number of
vertices and edges; i.e., in some contexts, $V = \norm{V}$ and $E = \norm{E}$. This abuse naturally extends to other
vertex/edge sets.)

Recall that $G$, by definition, contains no cycles or self-edges and its vertices can be topologically
ordered. We assume that $G$'s vertices have already been topologically ordered (in one of the possible ways) and labelled
from $1$ to $V$ (i.e., $V = \narset{1, \ldots, V}$) to reflect this topological order.\footnote{As famously known, this
order can be computed in $O(V+E)$ time \cite{cormen1990introduction}, which is faster than the algorithms we discuss in
this paper.} Consequently, there may exist a path from
vertex $i$ to vertex $j$ only if $i < j$.

Each edge in $E$ is a triplet $(i, j, w)$ where $1 \le i < j \le V$ and $w$ is a real number representing the
\dfn{weight}. There is at most one edge from a vertex $i$ to another vertex $j$.\footnote{There is no point to consider
multigraphs as only the shortest edge between two vertices matters.} We use the additional notation $w(i, j)$ to refer
to the weight of the edge from $i$ to $j$. If there is no edge from $i$ to $j$, then we assume $w(i, j) = \infty$. In
discussions where the weight is not significant, we simply use $(i, j)$ to refer to the edge from $i$ to $j$.

The \dfn{shortest path distance} (or shortly, the \dfn{distance}) from vertex $i$ to vertex $j$ is denoted by $d(i, j)$.
If there is no path from $i$ to $j$, we assume $d(i, j) = \infty$. Since there is no (negative) cycle in the graph, it
is not possible that $d(i, j) = -\infty$. Finally, $d(i, i) = 0$ for all vertices $i$.

In the \dfn{all-pairs shortest path problem} (for DAGs), shortly referred as \APSP, we want to compute the distances
$d(i, j)$ for all pairs of vertices $(i, j)$ in $G$.

The directional neighborhood of the vertices are of particular interest in our analyses. For a vertex $u$, $\INv{u}$
refers to $u$'s neighbors from which its incoming edges arrive, i.e., $\INv{u} = \set{i \such w(i, u) \not = \infty}$.
Similarly, $\OUTv{u}$ refers to the neighbors on $u$'s outgoing edges.

A fundamental concept in our algorithms is the partial graph. $G$'s \dfn{partial graph} from $i$
to $j$ (where $1 \le i \le j \le V$), denoted $\Gij{i}{j}$, is the graph restricted to the range of vertices $\narset{i,
\ldots, j}$. We refer to the vertices and edges of $\Gij{i}{j}$ with similar notation; in particular, $\Gij{i}{j} =
(\Vij{i}{j}, \Eij{i}{j})$ where $\Vij{i}{j} = \narset{i, \ldots, j}$ and $\Eij{i}{j} = \set{(u, v, w) \in E \such i \le u,
v \le j}$. The notation further extends to directional neighborhoods. For example, $\OUTijv{i}{j}{u}$ refers to $u$'s
neighbors on its outgoing edges restricted to the partial graph $\Gij{i}{j}$.

	\subsection{Essential Subgraph}
	To end up with specialized time bounds, we use the concept of an essential subgraph \cite{RobertEssential,
mcgeoch1995all}. An \dfn{essential subgraph} of $G$ is a graph on the same vertex set $V$ containing a minimal subset
of edges from $E$ such that the original distance function $d()$ is retained. The subgraph is minimal in the sense that,
if you were to remove any edge from it, the distance function $d()$ would be disturbed, i.e., some $d(i, j)$ would
increase. We use the notation $\Gess$ to refer to an essential subgraph of $G$ and $\Eess$ to refer to its edges, i.e.,
$\Gess = (V, \Eess)$.

To identify characterize the edges in $\Eess$, we use the
concept of an \dfn{essential edge}. For a DAG $G = (V, E)$ with the distance function $d()$, we say that an edge $(i, j,
w)$ is essential if $w < d(i, k) + d(k, j)$ for all $k \in \set{i + 1, \ldots, j - 1}$, i.e., there is no other route
of smaller or \fcs{equal} length from $i$ to $j$ than the direct edge itself. Notice that the
definition of an essential edge applies in the general graph $G$ and is not obviously linked to a minimal set of edges
that make up an essential subgraph. The following two lemmas imply that the essential subgraph is simply the subgraph of
essential edges and thus is unique, a fact previously argued by \cite{RobertEssential} for a larger class of graphs in
the more abstract framework of Q-semirings. Here, we provide simpler proofs specifically for real-weighted
DAGs.

\begin{lemma}\label{lem:essedges-are-in-essgraph}
    If $\Gess = (V, \Eess)$ is an essential subgraph of $G$, then every essential edge of $G$ is in $\Eess$.
\end{lemma}
\begin{proof}
    Assume to the contrary that an essential edge $(i, j, w) \not \in \Eess$.  By definition of essential edges, the
    shortest alternative path from $i$ to $j$ has a length $> w$. It follows that the distance $d(i, j)$ is larger in
    $\Gess$ than in $G$, a contradiction.
\end{proof}

\begin{lemma}\label{lem:essgraph-contains-essedges}
    If $\Gess = (V, \Eess)$ is an essential subgraph of $G$, then every edge of $\Eess$ is essential in $G$.
\end{lemma}
\begin{proof}
    Assume to the contrary that a non-essential edge $e = (i, j, w)$ of $G$ is in $\Eess$. Since $\Gess$ retains the
    original distance function $d()$ and $e$ is non-essential, there exists a $k \in \set{i + 1, \ldots, j - 1}$ such
    that $d(i, k) + d(k, j) \le w$ in $\Gess$, i.e., $e$ has an alternative route of smaller or equal length.
    It follows removing $e$ from $\Gess$ would still retain the distance function, contradicting that
    $\Gess$ is minimal.
\end{proof}

\begin{corollary}\label{cor:essential}
    The essential subgraph $\Gess$ of $G$ is unique and its edges are the essential edges of $G$.
\end{corollary}

Interestingly, the essential subgraph is unique not only for its host graph but also for its distance function, as shown
in the following lemma.

\begin{lemma}
    If $G_1$ and $G_2$ are two DAGs with the same distance function $d()$, then their essential subgraphs $\Gess_1$ and
    $\Gess_2$ are the same. In other words, there is a unique essential subgraph for a particular distance
    function.
\end{lemma}
\begin{proof}
    Since essential edges have no alternative shortest paths, for any essential edge $(i, j, w)$ either in $G_1$ or
    $G_2$, $w = d(i, j)$. If we construct another DAG $G$ that uses the distance function as its weights (i.e., $w(i, j)
    = d(i, j)$), then the essential edges of both $G_1$ and $G_2$ would be part of $G$; in other words, $\Gess_1$ and
    $\Gess_2$ are embedded in $G$. It follows that $\Gess_1$ and $\Gess_2$ each is an essential subgprah of $G$, which
    is unique, and thus $\Gess_1 = \Gess_2$.
\end{proof}

\paragraph{A note on the subgraph of optimal edges.}
Let us briefly note that another similar subgraph concept was also studied
in the literature \cite{karger1993finding, chung2007essential}, to which \cite{karger1993finding} refers as the subgraph
of ``optimal
edges''.\footnote{Interestingly, \cite{chung2007essential} reuses the term ``essential subgraph''
for this distinct concept.} We point out that, despite having a similar definition, this concept is distinct from the
essential subgraph in a fundamental way: The optimal edges are a maximal set of edges which take part in shortest paths,
whereas the essential subgraph is a minimal set of edges that keeps the shortest path distances intact. Consequently,
the essential subgraph may be significantly smaller and is more reasonable to use to argue specialized bounds.

	\subsection{Algorithm}
	\label{sec:algorithm}
	We now describe how to solve \APSP on DAGs within a specialized time bound based on the essential subgraph. Our
algorithm can be viewed as an extension to an algorithm by Chung \cite{chung2007essential}, which solves the same
problem by utilizing the subgraph of optimal edges.\footnote{See the note in the previous subsection on the subgraph of
optimal edges.} We modify Chung's approach to use the essential subgraph. Furthermore, we extend it to provide a more
refined time bound that involves a covariance depending (indirectly) on vertex degrees. We first explain the algorithm
and argue a basic time bound, followed by the more refined bound in the next subsection.

At the high level, our algorithm computes the distances and the essential edges inductively, in bottom-up fashion. In
particular, let the \dfn{range} of a pair of vertices $(i, j)$ (with $i < j$) be $j - i$. Due to the topological order,
any proper subpath of a path from $i$ to $j$ goes between a vertex pair $(u, v)$ of a smaller range.
Consequently, to build the shortest paths inductively, we consider all vertex pairs $(i, j)$ in increasing range order.
For each $(i, j)$, we compute the distance $d(i, j)$ and whether the edge $(i, j)$ is essential by using the
previously computed distances and essential edges on smaller ranges.

\begin{figure}[t]
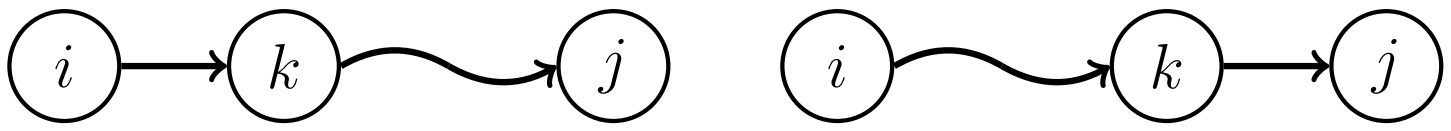

    \centering
    \subfloat[]{%
        \begin{graph}
            \node[vertex] (i) {$i$};
            \node[vertex, right of = i, node distance=8ex] (k) {$k$};
            \node[point, right of = k, node distance=6ex] (mid) {};
            \node[vertex, right of = mid, node distance=6ex] (j) {$j$};
            \path[->] (i) edge (k);
            \path[-] (k.0) edge[bend left=30] (mid);
            \path[->] (mid) edge[bend right=30] (j.180);
        \end{graph}%
    }%
    \hspace{4ex}%
    \subfloat[]{%
        \begin{graph}
            \node[vertex] (i) {$i$};
            \node[point, right of = i, node distance=6ex] (mid) {};
            \node[vertex, right of = mid, node distance=6ex] (k) {$k$};
            \node[vertex, right of = k, node distance=8ex] (j) {$j$};
            \path[-] (i.0) edge[bend left=30] (mid);
            \path[->] (mid) edge[bend right=30] (k.180);
            \path[->] (k) edge (j);
        \end{graph}%
    }
    \caption{The two ways of decomposing a shortest path from $i$ to $j$.}
    \label{fig:decomposition}
\end{figure}

The crux of the algorithm is how we compute each distance $d(i, j)$.
Notice that a shortest path from $i$ to $j$ can
be decomposed as
\fcs{both}:
\begin{enumerate}[label=(\alph*)]
    \item \label{dec:out} An essential edge from $i$ to some intermediate vertex $k$ followed by a shortest path from
          $k$ to $j$, and
    \item \label{dec:in} A shortest path from $i$ to some intermediate vertex $k$ followed by an essential edge from $k$
          to $j$
\end{enumerate}
unless it is the direct edge from $i$ to $j$. (See \autoref{fig:decomposition}.) To reduce the complexity (and for our
refined bound), we choose the decomposition that gives the smaller number of possibilities for $k$. In particular, if
the out-degree of $i$ is larger than the in-degree of $j$ (among the known essential edges), we choose the decomposition
\ref{dec:out}. Otherwise, \ref{dec:in} is chosen. By trying out every possible intermediate vertex $k$ for the chosen
decomposition, we compute the indirect shortest path distance between $i$ and $j$. If $w(i, j)$ is less than this
distance, then the edge $(i, j)$ is essential and $d(i, j) = w(i, j)$. Otherwise, the edge is non-essential and
$d(i, j)$ belongs to the indirect path.

Throughout the computation, we store the computed distances in a distance matrix $d$. The essential edges are stored in
a bidirectional adjacency list $\Eess$, on which we use $\INess/\OUTess$ to refer to the list of incoming/outgoing edges
of a particular vertex. \autoref{algo:apsp} provides the pseudocode for our approach. 

\begin{algorithm}[t]
    \begin{codebox}
        \zi $\Eess$ $\gets$ Empty bidirectional adjacency list on $V$
        \slit
        \zi $\dmat$ $\gets$ New $V\!\!\times\!V$ matrix
        \slit
        \zi Set $\dmat[i, i]$ $\gets$ $0$ for each $i \in \narset{1, \ldots, V}$.
        \slit
        \zi \For $r = 1$ to $V - 1$ \Do
        \zi     \For $i = 1$ to $V - r$ \Do
        \zi         $j \gets i + r$
        \zi         $t \gets \infty$
        \zi         \If $\norm{\OUTessv{i}} < \norm{\INessv{j}}$ \Then
        \zi             \For $k \in \OUTessv{i}$ \Do
        \zi                 $t \gets min(t, w(i, k) + \dmat[k, j])$ \End 
        \zi         \Else
        \zi             \For $k \in \INessv{j}$ \Do
        \zi                 $t \gets min(t, \dmat[i, k] + w(k, j))$ \End \End
        \zi         \If $w(i, j) < t$ \Then
        \zi             Insert $(i, j)$ into $\Eess$
        \slit
        \zi             $\dmat[i, j] \gets w(i, j)$
        \zi         \Else
        \zi             $\dmat[i, j] \gets t$ \End \End \End
    \end{codebox}
	\caption{\APSP on DAGs.}
	\label{algo:apsp}
\end{algorithm}

\begin{theorem}
    \APSP on DAGs can be solved in $O(V\!\Eess + V^2)$ time.
\end{theorem}
\begin{proof}
    By induction, our algorithm correctly computes the distance matrix (and the essential subgraph). The $V\!\Eess$ term
    in the bound is due to the inner loops on $k$: Each candidate vertex $k$ is uniquely associated with a particular
    essential edge (from $\OUTess(i)$ or $\INess(j)$) and a particular destination/source vertex against that edge ($j$
    or $i$). (Revisit \autoref{fig:decomposition} to visualize the association.) The additional term of $V^2$ is due to
    the outer loops.
\end{proof}

	\subsection{Refining the Bound}
	\label{sec:refinement}
	With a more precise
analysis, the $\basicterm$ term in the time bound can be improved. Let $\INess$/$\OUTess$ refer to the directional
neighborhoods in $\Gess$. Also, let $\Gessij{i}{j}$ be $\Gess$'s partial graph from $i$ to $j$ and let
$\INessij{i}{j}$/$\OUTessij{i}{j}$ refer to the directional neighborhoods in $\Gessij{i}{j}$. When the pair $(i, j)$ is
processed, since we consider pairs in increasing range, the discovered essential edges which extend from $i$ is exactly
$\OUTessijv{i}{(j-1)}{i}$. Similarly, the discovered incoming edges to $j$ is $\INessijv{(i+1)}{j}{j}$. We write the
following for the number of iterations in the inner loops:

\begin{align*}
          \sum_{1 \le i < j \le V} \minfun{\norm{\OUTessijv{i}{(j-1)}{i}}, \norm{\INessijv{(i+1)}{j}{j}}}
    \le & \sum_{1 \le i , j \le V} \minfun{\norm{\OUTessv{i}}, \norm{\INessv{j}}} \\[1ex]
    =   & \sum_{1 \le i , j \le V} \sum_{t=1}^{\minfun{\norm{\OUTessv{i}}, \norm{\INessv{j}}}} ~1 \\[1ex]
    =   & \sum_{1 \le t \le V} ~ \sum_{\substack{1 \le i \le V\\[0.5ex]\norm{\OUTessv{i}} \ge t}} ~
                                 \sum_{\substack{1 \le j \le V\\[0.5ex]\norm{\INessv{j}} \ge t}} ~1 \\[1ex]
    =   & \sum_{1 \le t \le V} \DistOUTt{t}\DistINt{t}
\end{align*}

where $\DistOUTt{t}$ ($\DistINt{t}$) is the number of vertices whose out-degree (in-degree) is at least $t$ in the
essential subgraph $\Gess$. In other words, $\DistOUTt$ and $\DistINt$ are \dfn{the outgoing and the incoming degree
exceedance distributions} of $\Gess$.
Now, observe that

\[ \sum_{1 \le t \le V}\DistOUTt{t} ~=~ \sum_{1 \le t \le V}\sum_{\substack{1 \le i \le V\\[0.5ex]\norm{\OUTessv{i}} \ge
t}} 1 ~=~ \sum_{1 \le i \le V} ~ \sum_{1 \le t \le \norm{\OUTessv{i}}} 1 ~=~ \Eess \]

and hence the mean of $\DistOUTt{t}$ (and similarly the mean $\DistINt{t}$) is $\Eess/V$, to which we refer as $\mu$.
(An important observation at this point is that $\mu$ is also the average outgoing/incoming degree in the graph.) We
write the factors in the previous bound as deviations from the mean (with $\DeltaOUTt{t} = \DistOUTt{t} - \mu$ and
$\DeltaINt{t} = \DistINt{t} - \mu$):

\begin{align*}
    \hspace{4.5ex}&\hspace{-4.5ex}\sum_{1 \le t \le V} \DistOUTt{t}\DistINt{t}
    = \sum_{1 \le t \le V} (\mu + \DeltaOUTt{t})(\mu + \DeltaINt{t}) \\[1ex]
    & = \sum_{1 \le t \le V} \mu^2 + \mu\DeltaINt{t} + \mu\DeltaOUTt{t} + \DeltaOUTt{t}\DeltaINt{t} \\[1ex]
    & = \sum_{1 \le t \le V}\mu^2 + \sum_{1 \le t \le V}\mu\DeltaINt{t} + \sum_{1 \le t \le V}\mu\DeltaOUTt{t} + \sum_{1 \le t \le V}\DeltaOUTt{t}\DeltaINt{t} \\[1ex]
    & = V\mu^2 + \sum_{1 \le t \le V}\DeltaOUTt{t}\DeltaINt{t} \\[1ex]
\end{align*}
Here, the first term equals $\refinedterm$. The second term involves products of differences from the mean and is
therefore $V$ times the \dfn{covariance} of the exceedance distributions $\DistOUT$ and
$\DistIN$. We denote this covariance by $\sigma$. Hence, $O\left(\refinedterm + \covariantterm + V^2\right)$ is another
time bound for our algorithm.

\begin{theorem}\label{thm:refinement}
    \APSP for DAGs can be solved in $O\left(\refinedterm + \covariantterm + V^2\right)$ time.
\end{theorem}

It is not immediately clear whether this new bound is better. To begin with, the following lemma shows that it is not
worse.

\begin{lemma}\label{lem:refinement-is-not-worse}
    For any DAG $G$, $\refinedterm + \covariantterm \le \basicterm$.
\end{lemma}
\begin{proof}
    Going back to the original expression that yields $\refinedterm + \covariantterm$:
    \begin{align*}
        \refinedterm + \covariantterm
        = & \sum_{1 \le i, j \le V} \minfun{\norm{\OUTessv{i}}, \norm{\INessv{j}}} \\
        = & \sum_{1 \le i, j \le V} \frac{\norm{\OUTessv{i}} + \norm{\INessv{j}}}{2} - \frac{\big|\!\norm{\OUTessv{i}} - \norm{\INessv{j}}\!\big|}{2} \\
        = & ~\basicterm - \sum_{1 \le i, j \le V} \frac{\big|\!\norm{\OUTessv{i}} - \norm{\INessv{j}}\!\big|}{2} \le \basicterm
    \end{align*}
\end{proof}

More significantly, the new bound can be asymptotically better for some classes of graphs, as implied by the following
lemma.

\begin{lemma}\label{lem:refinement-is-better}
    For any given number of vertices $V$, one can construct a DAG $G$ for which $\refinedterm + \covariantterm$ is
    $o(\basicterm)$ (as $V$ grows).
\end{lemma}
\begin{proof}
    For simplicity, we construct $G$ as a graph with unit weights and thus all edges are essential, i.e., $E = \Eess$.
    The main idea is to bound $\sigma$ by bounding the individual variances of the out-degree/in-degree exceedance
    distributions. In particular, let $\sigma_{\OUT}^2$ ($\sigma_{\IN}^2$) be the variance of the
    out-degree (in-degree) exceedance distribution $\DistOUT$ ($\DistIN$). Recall that the means of $\DistOUT$ and
    $\DistIN$ are both $\mu = E/V$, which is the average (directional) degree in the graph. Since both distributions are
    bounded by $V$, the Bhatia-Davis inequality implies that $\sigma_{\OUT}^2$ and $\sigma_{\IN}^2$ are both $O(\muV) =
    O(E)$. Also, from the Cauchy-Schwarz inequality, we know that $\sigma \le \sqrt{\sigma_{\OUT}^2\,\sigma_{\IN}^2}$.
    Thus, if we construct $G$ such that either $\sigma_{\OUT}^2$ or $\sigma_{\IN}^2$ is reduced to $o(E)$ and $E =
    o(V^2)$, then it follows that $\refinedtermE + \covariantterm = o(\basictermE)$.

    We now describe how to construct a DAG $G$ so that $\sigma_{\OUT}^2 = o(E)$. Select a number of edges $E$ such that
    $V \ll E \ll V^2$ (where $\ll$ denotes a strict asymptotic order). Then, the average degree $\mu = E/V$ satisfies $1
    \ll \mu \ll V$. We create $G$ in three parts, namely A, B and Z, each populated with arbitrary vertices
    and outgoing edges to yield the following characteristics:

    \begin{tabularx}{\textwidth}{@{\strut}l@{~}X@{\strut}}
        \textbf{Part A:} & Has $V_A$ vertices, each with an out-degree of $d_A$. \\
        \textbf{Part B:} & Has $V_B$ vertices, each with an out-degree of $d_B$. \\
        \textbf{Part Z:} & Has $d_A + d_B$ vertices, each with zero out-degree. All vertices in this part are
                           topologically higher than those in A and B, so that we can populate the desired
                           outgoing edges to yield a consistent DAG.
    \end{tabularx}

    To support further argument, the parameters are selected as follows:
    \begin{itemize}
        \item We choose any $d_A$ and $V_B$ such that $d_A \ll \mu \ll V_B \ll V$.
        \item We solve the linear equations $V = V_A + V_B + d_A + d_B$ and $E = V_Ad_A + V_Bd_B$ to determine $V_A$ and
              $d_B$. (Both $V_A$ and $d_B$ solve to positive values. However, they may solve to decimals, a situation
              that we address later.)
    \end{itemize}
    Observe that $d_B = (E - V_Ad_A)/V_B \le \muV/V_B = o(V)$ and thus $V_A = \Theta(V)$. Also, $d_B > d_A$ since $d_A <
    \mu$.

    The out-degree exceedance distribution $\DistOUT$ now contains the following pieces, each of which contributes to
    the variance $\sigma_{\OUT}^2$ as follows:
    \begin{itemize}
        \item For $t = 1 \ldots d_A$, $\DistOUTt{t} = V_A + V_B$. The variance contribution is
              \begin{center}$d_A(V_A + V_B - \mu)^2/V = o(\muV)$\end{center}
        \item For $t = (d_A + 1) \ldots d_B$, $\DistOUTt{t} = V_B$. The variance contribution is
              \begin{center}$(d_B - d_A)(V_B - \mu)^2/V \le {d_B}{V_B^2}/V \le {\muV}V_B/V = o(\muV)$\end{center}
        \item For $t = (d_B + 1) \ldots V$, $\DistOUTt{t} = 0$. The variance contribution is
              \begin{center}$(V - d_B)\mu^2/V = o(\muV)$\end{center}
    \end{itemize}
    Since $E = \muV$, it follows that $\sigma_{\OUT}^2 = o(E)$, essentially completing the proof.

    We finally mention how to deal with the case where $V_A$ and $d_B$ solve to decimals. In such cases, we simply use
    $\lfloor{V_A}\rfloor$ and $\lfloor{d_B}\rfloor$ in the construction. However, we will be two vertices short of $V$
    and $(d_A + V_B)$ edges short of $E$. This shortage can be easily compensated by prepending two dummy vertices and
    adding edges from them. The compensation does not change the asymptotic properties of our parameters and its impact
    on the variance is $o(V)$ and insignificant.
\end{proof}

We point out that the construction in our proof is significantly flexible and valid for any $E$ satisfying $V \ll E \ll
V^2$, implying that the refined bound is better a large variety of DAGs.

    \subsection{Transitive Closure and Reduction}
	\label{sec:transitive}
    Our \APSP algorithm can be easily adapted to find the transitive closure of a DAG. Recall that, given a DAG (or more
generally a digraph) $G = (V, E)$, the \dfn{transitive closure} problem asks to compute a new set $\Eclo$ of unweighted
edges on $V$ such that the edge $(u, v) \in \Eclo$ if and only if there exists a path from $u$ to $v$ in the original
graph. In other words, $\Eclo$ tells whether a given vertex $v$ is reachable from another given vertex $u$.

In the context of transitive closure, the original edge weights do not matter. We simply run our algorithm on the
unweighted form of $E$ and compute the transitive connections rather than the shortest path distances. This standard
adaptation famously involves defining a Boolean weight function  $w(u, v) = (u, v) \in E$ and replacing the semiring
$(\Reals, +, \min)$ with $(\narset{\True, \False}, \land, \lor)$, resulting in new inner lines in \autoref{algo:apsp}
(such as $t \gets t \lor (w(i, k) \land \dmat[k, j]$). The output matrix $d$ then contains a $\True$ (true) entry for
every edge of $\Eclo$.

The essential subpgraph in this context (with the new Boolean weight function) is potentially smaller than that for the
original weighted graph. It consists of a minimal set of edges that keeps the transitive connectivity between the
vertices intact. This set of edges is traditionally called the \dfn{transitive reduction} of $G$, which we denote by
$\Ered$ (as done in \cite{simon1988improved}).\footnote{Since
\Cref{lem:essedges-are-in-essgraph,lem:essgraph-contains-essedges} can be easily adapted to apply to $\Ered$, $\Ered$ is
unique and consists of essential edges without alternative paths. Another intuitive view of $\Ered$ is that it is the
set of essential edges when all edge weights are fixed as $1$.} Since our adapted algorithm also computes the essential
edges (i.e., the transitive reduction), we can adapt \Cref{thm:refinement} as follows.

\begin{theorem}\label{thm:transitive}
    The transitive closure and the transitive reduction for a given DAG $G$ can be computed in $O\left(\refinedtermEred
    + \covarianttermred + V^2\right)$ time, where $\sigmared$ is the covariance of the in-degree/out-degree exceedance
    distributions restricted to the edges of $\Ered$.
\end{theorem}

    \section{All-Pairs Lowest Common Ancestor in DAGs}
    \label{sec:aplca}

    This section covers the all-pairs lowest common ancestor algorithm for DAGs. This algorithm not only computes a
    lowest common ancestor but also the deepest common ancestor between each pair of vertices. Similar to the \APSP
    algorithm, the time bound involves a parameter, based on the transitive reduction graph. As we shall explain, the
    parameter, this time, can be written in terms of the variances of the in-degree and the out-degree exceedance
    distributions. We begin the explanation with the preliminaries.

    \subsection{Preliminaries}
    Consider an unweighted DAG $G = (V, E)$, where $V$ is the set of vertices and $E$ is the set of directed edges. As in
\Cref{sec:preliminaries}, the vertices in $V$ are already topologically ordered and labelled from $1$ to $V$. (We again
abuse $V$ and $E$ to refer to $\norm{V}$ and $\norm{E}$.) We say that a vertex $i$ is a \dfn{predecessor} of a vertex
$j$ if the directed edge $(i, j)$ exists. In reverse, $j$ is said to be a \dfn{successor} of $i$. Extending this
relationship transitively, we say that a vertex $i$ is an \dfn{ancestor} of another vertex $j$ if there
exists a path from $i$ to $j$. In reverse, $j$ is said to be a \dfn{descendant} of $i$. Clearly, $i \le j$ if $i$ is an
ancestor of $j$, but the converse is not true in general. Every vertex is its own ancestor and descendant.

A vertex $v$ is a \dfn{lowest common ancestor}, shortly an LCA, of two vertices $i$ and $j$, if $a$ is an ancestor of
both $i$ and $j$, and no successor of $v$ (or equivalently no descendant of $v$) is an ancestor of both $i$ and $j$. It
is well-known that if the DAG mimics a rooted tree, then the LCA for any pair of vertices uniquely exists. For general
DAGs, however, it is possible the LCA does not exist or there are multiple LCAs. $\LCAij{i}{j}$ refers to the set of
LCAs for any pair $i, j$.

In the \dfn{all-pairs lowest common ancestor problem} (for DAGs), shortly referred as \APLCA, we want to compute a
\fcs{single} LCA vertex (if any) for each pair of vertices $(i, j)$ in $G$.

A concept central to our approach is the vertex depth. The \dfn{depth} of a vertex $i$, denoted $\depthi{i}$, is the
length of the longest path (in terms of the number of nodes) that ends in $i$. Naturally, the depth of any vertex is
inclusively between $1$ and $V$.

A vertex $v$ is called a \dfn{deepest common ancestor}, shortly a DCA, of two vertices $i$ and $j$, if $v$ is a
common ancestor of $i$ and $j$ and has maximum depth among such common ancestors. Observe that, since a successor of any
vertex has a higher depth, the successor of a DCA cannot be a common ancestor, and thus every DCA is an LCA. The converse,
as expected, is not true in general. Two vertices may have zero, one, or multiple DCAs; $\DCAij{i}{j}$ refers to the set
of DCAs for any pair $i, j$. As an abuse of notation, $\DCAij{i}{j}$ also refers to the depth of vertices in
$\DCAij{i}{j}$. Finally, observe that $\LCAij{i}{j} = \emptyset$ if and only if $\DCAij{i}{j} = \emptyset$.

    \subsection{Algorithm}
    We solve the \APLCA problem by solving a specialization; we compute the DCA for all pairs. (Recall that every DCA
is an LCA.) The use of the DCA allows us to write a particular recurrence relation as to be exhibited soon. To have a
base case for this relation and to bound our running time, our algorithm starts by computing the transitive closure
$\Eclo$ and the transitive reduction $\Ered$ of $G$ (as explained in \Cref{sec:transitive}).

Observe that replacing $E$ with $\Ered$ would have no effect on the ancestor/descendant relationships since the
original transitive closure $\Eclo$ would still be intact. Interestingly, neither does it have any effect on the depths
of vertices. This happens in particular because the reduction would only eliminate the edges with alternative longer
paths, and thus the longest path to any vertex is still intact. In other words, we can compute the LCAs or the DCAs on
the reduced graph $\Gred = (V, \Ered)$, since they retain their LCA/DCA relationships.

We can then write the following recurrence for $\DCAij{i}{j}$ (where $\DCAij{i}{j}$ refers to the depth of the
DCAs and $\max$ resolves to $0$ for empty sets):
\begin{flalign*}
    \DCAij{i}{j} = \begin{cases}
                       \depthi{i}                                         & \mbox{if }(i, j) \in \Eclo \\
                       \depthi{j}                                         & \mbox{if }(j, i) \in \Eclo \\
                       \displaystyle\max_{\mbox{\small\vphantom{İ}$(k, i) \in \Ered$}} \DCAij{k}{j}  & \mbox{otherwise}
                   \end{cases}
\end{flalign*}
Let us clarify the recurrence. If either $i$ or $j$ is an ancestor of the other, then it is clearly a DCA. (The first
two cases of the recurrence.) Otherwise, a DCA of $i$ and $j$ is also a DCA of some predecessor $k$ of $i$ and $j$.
Thus, if we look at all predecessors $k$ of $i$ and take deepest DCA they make with $k$, we find a DCA of $i$ and $j$.

Notice that the same recurrence could be in an alternative form, branching on the predecessors of $j$ instead. Thus, we
get two choices for the recurrence. As done previously in \Cref{sec:algorithm}, we can choose the one with the smaller
number of options to argue a specialized bound.

Our algorithm now is simply a bottom-up dynamic programming of this recurrence. After precomputing the depths of all
vertices\footnote{Notice that we can compute the depths by going through the vertices in topological order in $O(V +
\Ered)$ time. The depth of a particular vertex is simply one plus the depth of its deepest predecessor.}, we visit all
pairs $(i, j)$ in dependency order of the recurrence (e.g., we visit them in increasing lexicographical order) and
populate a memo table storing the solutions to the subproblems. While constructing the solutions, we apply the form of
the recurrence that minimizes the number of branches. In particular, let $\INredv{v}$ denote the predecessors of $v$ in
the reduced graph. If $\norm{\INredv{i}} < \norm{\INredv{j}}$, the above form of the recurrence is used. Otherwise, the
alternative form is used. \autoref{algo:aplca} details our approach.

\begin{algorithm}[t]
    \begin{codebox}
        \zi $\Eclo, \Ered$ $\gets$ Compute the transitive closure and reduction of $G$.
        \slit
        \zi $\depth$ $\gets$ Compute the depths of all vertices in $G$. (Also, $\depth[\emptyset] = 0$.)
        \slit
        \zi $\DCA$ $\gets$ New empty $V\!\!\times\!V$ table.
        \slit
        \zi \For $i = 1$ to $V$ \Do
        \zi     \For $j = i$ to $V$ \Do
        \zi         \If $(i, j) \in \Eclo$ \Then
        \zi             $\DCA[i, j] \gets \DCA[j, i] \gets i$
        \zi         \Else
        \zi             $t \gets \emptyset$
        \zi             \If $\norm{\INredv{i}} < \norm{\INredv{j}}$ \Then
        \zi                 \For $k \in \INredv{i}$ \Do
        \zi                     \If $\depth[t] < \depth[\DCA[k, j]]$ \Then
        \zi                         $t \gets \DCA[k, j]$ \End \End
        \zi             \Else
        \zi                 \For $k \in \INredv{j}$ \Do
        \zi                     \If $\depth[t] < \depth[\DCA[i, k]]$ \Then
        \zi                         $t \gets \DCA[i, k]$ \End \End \End
        \zi             $\DCA[i, j] \gets \DCA[j, i] \gets t$ \End \End \End
    \end{codebox}
	\caption{\APLCA on DAGs.}
	\label{algo:aplca}
\end{algorithm}

    \subsection{Parameterized Bound}
    Similar to the \APSP algorithm, our \APLCA algorithm also gets a parameterized bound. However, this time, it primarily
involves a variance rather than a covariance, as we are about to explain.

To begin with, we analyze the total cost of the innermost loops:

\begin{align*}
          \sum_{1 \le i \le j \le V} \minfun{\norm{\INredv{i}}, \norm{\INredv{j}}}
    \le & \sum_{1 \le i, j \le V} \minfun{\norm{\INredv{i}}, \norm{\INredv{j}}}
\end{align*}

By following similar steps to those in \Cref{sec:refinement}, this expression equals:

\begin{align*}
    \sum_{1 \le t \le V} \DistINredt{t}^2
\end{align*}

where $\DistINred$ is the in-degree exceedance distribution of $\Gred$. Notice that, since the two recurrences we
utilize both use incoming edges (rather than incoming versus outgoing), we get squares of the in-degree exceedances
as the terms. The mean of $\DistINred$ is $\mu = \Ered/V$. Again, by going through steps as in \Cref{sec:refinement},
we transform the above sum to:

\begin{align*}
    \Ered^2/V + \sum_{1 \le t \le V} \DeltaINredt{t}^2
\end{align*}

where $\DeltaINredt{t} = \DistINredt{t} - \mu$. Here, the second term is the variance of in-degree exceedance
distribution $\DistINred$, which we denote by $\sigmasquareINred$. Therefore, $O(\Ered^2/V + \variantterm)$ is an
upper bound for the cost of inner loops.

The cost for the inner loops are accompanied by an $O(\quadraticterm)$ cost for the outer loops, an $O(V + \Ered)$ cost
for computing the depths and an $O(\refinedtermEred + \covarianttermred + \quadraticterm)$ cost for the transitive
closure/reduction (where $\sigma$ is the covariance of $\DistINred$ and $\DistOUTred$) by \Cref{thm:transitive}.
Summing all the costs, we arrive at the following theorem.

\begin{theorem}
    \APLCA for DAGs can be solved in $O(\refinedtermEred + \variantmixterm + \quadraticterm)$ time.
\end{theorem}

Relying on the Cauchy-Schwarz inequality $\sigma \le \sqrt{\sigmasquareINred\,\sigmasquareOUTred}$,
we can write the $\variantmixterm$ term more elegantly as $\variantmaxterm$, at the expense of slightly
increasing the bound.\footnote{Here, $\sigmasquareOUTred$ is the variance of the out-degree exceedances in the
transitive reduction.} Let us also mention that the bound would be better in the cases where the transitive
closure/reduction is performed faster, e.g., by utilizing an algorithm sensitive to the chains \cite{simon1988improved}.

	\bibliographystyle{unsrt} 
	\bibliography{references}  

\begin{thebibliography}{10}

\bibitem{mcgeoch1995all}
Catherine~C. McGeoch.
\newblock All-pairs shortest paths and the essential subgraph.
\newblock {\em Algorithmica}, 13(5):426--441, 1995.

\bibitem{goralvcikova1979reduct}
Alla Goral{\v{c}}{\'\i}kov{\'a} and V{\'a}clav Koubek.
\newblock A reduct-and-closure algorithm for graphs.
\newblock In {\em International Symposium on Mathematical Foundations of
  Computer Science}, pages 301--307. Springer, 1979.

\bibitem{chung2007essential}
Eesuk Chung.
\newblock Essential arcs of a directed acyclic graph.
\newblock {\em Management Science and Financial Engineering}, 13(1):121--126,
  2007.

\bibitem{ford1956network}
Lester~Randolph Ford.
\newblock Network flow theory.
\newblock {\em Rand Corporation Paper, Santa Monica, 1956}, 1956.

\bibitem{bellman1958routing}
Richard Bellman.
\newblock On a routing problem.
\newblock {\em Quarterly of applied mathematics}, 16(1):87--90, 1958.

\bibitem{Dijkstra1959}
E.~W. Dijkstra.
\newblock A note on two problems in connexion with graphs.
\newblock {\em Numerische Mathematik}, 1(1):269–271, December 1959.

\bibitem{fredman1987fibonacci}
Michael~L Fredman and Robert~Endre Tarjan.
\newblock Fibonacci heaps and their uses in improved network optimization
  algorithms.
\newblock {\em Journal of the ACM (JACM)}, 34(3):596--615, 1987.

\bibitem{warshall1962theorem}
Stephen Warshall.
\newblock A theorem on boolean matrices.
\newblock {\em Journal of the ACM (JACM)}, 9(1):11--12, 1962.

\bibitem{madkour2017survey}
Amgad Madkour, Walid~G Aref, Faizan~Ur Rehman, Mohamed~Abdur Rahman, and Saleh
  Basalamah.
\newblock A survey of shortest-path algorithms.
\newblock {\em arXiv preprint arXiv:1705.02044}, 2017.

\bibitem{williams2014faster}
Ryan Williams.
\newblock Faster all-pairs shortest paths via circuit complexity.
\newblock In {\em Proceedings of the forty-sixth annual ACM symposium on Theory
  of computing}, pages 664--673, 2014.

\bibitem{pettie2004new}
Seth Pettie.
\newblock A new approach to all-pairs shortest paths on real-weighted graphs.
\newblock {\em Theoretical Computer Science}, 312(1):47--74, 2004.

\bibitem{karger1993finding}
David~R Karger, Daphne Koller, and Steven~J Phillips.
\newblock Finding the hidden path: Time bounds for all-pairs shortest paths.
\newblock {\em SIAM Journal on Computing}, 22(6):1199--1217, 1993.

\bibitem{fischer1971boolean}
Michael~J Fischer and Albert~R Meyer.
\newblock Boolean matrix multiplication and transitive closure.
\newblock In {\em 12th annual symposium on switching and automata theory (swat
  1971)}, pages 129--131. IEEE, 1971.

\bibitem{aho1972transitive}
Alfred~V. Aho, Michael~R Garey, and Jeffrey~D. Ullman.
\newblock The transitive reduction of a directed graph.
\newblock {\em SIAM Journal on Computing}, 1(2):131--137, 1972.

\bibitem{simon1988improved}
Klaus Simon.
\newblock An improved algorithm for transitive closure on acyclic digraphs.
\newblock {\em Theoretical Computer Science}, 58(1-3):325--346, 1988.

\bibitem{kogan2022beating}
Shimon Kogan and Merav Parter.
\newblock Beating matrix multiplication for n\^{}$\{$1/3$\}$-directed
  shortcuts.
\newblock In {\em 49th International Colloquium on Automata, Languages, and
  Programming (ICALP 2022)}, pages 82--1. Schloss Dagstuhl--Leibniz-Zentrum
  f{\"u}r Informatik, 2022.

\bibitem{czumaj2007faster}
Artur Czumaj, Miros{\l}aw Kowaluk, and Andrzej Lingas.
\newblock Faster algorithms for finding lowest common ancestors in directed
  acyclic graphs.
\newblock {\em Theoretical Computer Science}, 380(1-2):37--46, 2007.

\bibitem{blelloch2008new}
Guy~E Blelloch, Virginia Vassilevska, and Ryan Williams.
\newblock A new combinatorial approach for sparse graph problems.
\newblock In {\em International Colloquium on Automata, Languages, and
  Programming}, pages 108--120. Springer, 2008.

\bibitem{grandoni2021all}
Fabrizio Grandoni, Giuseppe~F Italian, Aleksander {\L}ukasiewicz, Nikos
  Parotsidis, and Przemys{\l}aw Uzna{\'n}ski.
\newblock All-pairs lca in dags: Breaking through the o (n 2.5) barrier.
\newblock In {\em Proceedings of the 2021 ACM-SIAM Symposium on Discrete
  Algorithms (SODA)}, pages 273--289. SIAM, 2021.

\bibitem{kowaluk2007unique}
Miros{\l}aw Kowaluk and Andrzej Lingas.
\newblock Unique lowest common ancestors in dags are almost as easy as matrix
  multiplication.
\newblock In {\em European Symposium on Algorithms}, pages 265--274. Springer,
  2007.

\bibitem{kowaluk2008path}
Miros{\l}aw Kowaluk, Andrzej Lingas, and Johannes Nowak.
\newblock A path cover technique for lcas in dags.
\newblock In {\em Scandinavian Workshop on Algorithm Theory}, pages 222--233.
  Springer, 2008.

\bibitem{eckhardt2007fast}
Stefan Eckhardt, Andreas~Michael M{\"u}hling, and Johannes Nowak.
\newblock Fast lowest common ancestor computations in dags.
\newblock In {\em European Symposium on Algorithms}, pages 705--716. Springer,
  2007.

\bibitem{cormen1990introduction}
Thomas~H Cormen, Charles~Eric Leiserson, Ronald~L Rivest, and Clifford Stein.
\newblock {\em Introduction to algorithms}, volume~2.
\newblock MIT press, Cambridge, 1990.

\bibitem{RobertEssential}
Pierre Robert.
\newblock An algorithm for finding the essential sets of arcs of certain
  graphs.
\newblock {\em Journal of Combinatorial Theory, Series B}, 10(3):288--298,
  1971.

\end{thebibliography}
	
\end{document}